\documentclass[apjl]{emulateapj}
\usepackage{amsfonts,amsmath,graphicx,natbib,apjfonts,subfigure}

\def\cfa{1}
\def\ferr{2}
\def\nasa{4}
\def\oregon{3}

\shorttitle{Super-soft late-time X-ray emission in GRBs}
\shortauthors{Margutti et al.}

\begin{document}
%\title{Playing con sordina: powerful Gamma-Ray Burst explosions shrouded in significant circum-burst material}
\title{Dust in the wind: the role of recent mass loss in Long Gamma-Ray Bursts}
%Dust in the wind
\author{R.~Margutti\altaffilmark{\cfa}, C. Guidorzi\altaffilmark{\ferr}, D. Lazzati\altaffilmark{\oregon}, 
D. Milisavljevic\altaffilmark{\cfa}, A. Kamble\altaffilmark{\cfa},  T. Laskar\altaffilmark{\cfa}, J. Parrent\altaffilmark{\cfa}, N.~C. Gehrels\altaffilmark{\nasa}, A.~M. Soderberg\altaffilmark{\cfa}}

\altaffiltext{\cfa}{Harvard-Smithsonian Center for Astrophysics, 60 Garden St., Cambridge, MA 02138, USA}
\altaffiltext{\ferr}{Department of Physics and Earth Sciences, University of Ferrara, via Saragat 1, I-44122, Ferrara, Italy}
\altaffiltext{\oregon}{Department of Physics, Oregon State University, 301 Weniger Hall, Corvallis, OR 97331, USA}
\altaffiltext{\nasa}{NASA Goddard Space Flight Center, Greenbelt, MD 20771, USA}
\begin{abstract}
We study the late-time ($t>0.5$ days) X-ray afterglows of nearby ($z<0.5$) long
Gamma-Ray Bursts (GRB) with \emph{Swift} and identify a population of explosions with slowly
decaying, super-soft  (photon index $\Gamma_x>3$) X-ray emission that is inconsistent
with forward shock synchrotron radiation associated with the afterglow. These explosions also show
larger-than-average intrinsic absorption ($NH_{x,i}>6\times 10^{21}\,\rm{cm^{-2}}$) and 
prompt $\gamma$-ray emission  with extremely long duration ($T_{\rm{90}}>1000$ s).
Chance association of these three rare properties  (i.e. large $NH_{x,i}$, super-soft $\Gamma_x$ 
and extreme duration) in the same class of explosions is statistically unlikely. 
We associate these properties with the turbulent mass-loss history of the progenitor
star that enriched and shaped the circum-burst medium.  
We identify a natural connection between $NH_{x,i}$, $\Gamma_x$ and $T_{\rm{90}}$
in these sources by suggesting that the late-time super-soft X-rays originate from radiation 
reprocessed by material lost to the environment by the stellar progenitor before exploding, 
(either in the form of a dust echo 
or as reprocessed radiation from a long-lived GRB remnant), and that the
interaction of the explosion's shock/jet with the complex medium is the source
of the extremely long prompt emission. However, current observations do not allow us to exclude the possibility 
that super-soft X-ray emitters originate from peculiar stellar
progenitors with large radii that only form in very dusty environments.
%We suggest that the late-time super-soft X-rays originate from radiation 
%reprocessed by material lost to the environment by the stellar progenitor before exploding, 
%(either in the form of a dust echo or Compton echo of the prompt photons 
%or as reprocessed radiation from a long-lived GRB remnant), and that the
%interaction of the explosion's shock/jet with the complex medium might be at the origin
%of the extremely long prompt emission,
%thus identifying a natural connection between $NH_{x,i}$, $\Gamma_x$ and $T_{\rm{90}}$
%in these sources.
%We suggest that .... physical link.  Our results instead suggest a physical link. We envision 
%two possible scenarios. We speculate that.
\end{abstract}

%%%%%%%%%%%%%%%%%%%%%%%%%%%%%%%%%%%%%%%%%
\section{Introduction}
\label{Sec:Intro}

The effects of binarity and the role of mass loss in the decades to years preceding the
terminal explosion are among the least understood aspects of massive stellar evolution
 (e.g. \citealt{Langer12}, \citealt{Smith14} for recent reviews). 
This lack of understanding is significant in light of
 recent observations showing that more than $70\%$ of massive O-type stars 
in the Galaxy interact with a binary companion \citep{Sana12} and that the classical picture of
mass loss through steady winds does \emph{not} apply to all massive stars,
%and might not apply to evolved massive stars as a whole
especially during the very last stages of evolution  (e.g. \citealt{Ofek13}, \citealt{Margutti14} and references therein).

Long Gamma-ray Bursts (GRBs) are thought to represent the endpoints
of the evolution of massive stars that managed to lose their hydrogen 
envelope before exploding, while retaining enough angular momentum 
to power a relativistic jet (e.g. \citealt{MacFadyen99}, \citealt{MacFadyen01}). 
The observed connection of long GRBs with Type Ic supernovae (SNe) and their locations
within the host galaxies (\citealt{Fruchter06}) strongly support their association with massive         
stars and are consistent with the suggested Wolf-Rayet (WR) progenitors  (see
\citealt{Hjorth12} for a recent review).
However, it is  unclear if the progenitors of GRBs are single massive 
stars or binaries. The rate and the nature of the mass loss 
(e.g. steady winds vs. explosive ejection of shells of material) suffered by the stellar
progenitor in the final phases of its evolution before collapsing are also unclear.
 
Observations are now starting to reveal the turbulent life of some massive
stars in the years before the SN explosion and are pointing to the presence of
a common (and unexpected) eruptive behavior preceding the collapse (e.g. \citealt{Ofek14},
\citealt{Smith14}).
As a result, the local SN environment is shaped and ``enriched'' by successive mass ejections.
This kind of behavior might be common in GRB progenitor stars as well.
In particular, it is relevant to mention 
%that might be revealed as the SN blast wave and/or the jet 
%What is relevant to GRB science is 
(i) the recent report of a possible outburst of the
progenitor of the Type Ic SN PTF11qcj  $\sim2\,\rm{yrs}$ before the SN \citep{Corsi14},
(ii) the signature of increased mass loss shortly before the explosion 
of the hydrogen-stripped SNe 2013cu \citep{GalYam14} and 2008D \citep{Svirski14},
(iii) the finding of unusual environments around some Type Ib/c SNe as revealed
by radio observations (e.g. \citealt{Berger03a}, \citealt{Soderberg06b}, \citealt{Soderberg06d}, \citealt{Wellons12},
citealt{Bietenholz14}).

It is thus likely that a complex environment sculpted by 
the recent mass loss of the progenitor system
surrounds GRBs at the time of their explosions. The interaction of the 
GRB jet and the SN ejecta with this material is expected to leave detectable signatures
in their temporal and spectral evolution.
Here we present a study of GRBs in the low-redshift Universe ($z<0.5$)
that aims to test this hypothesis. 
We identify a class of explosions with peculiar prompt $\gamma$-ray emission and
late-time X-ray spectrum and connect these properties with the
mass-loss history of their progenitors.

Throughout the paper we use the convention $F_{\nu}(\nu,t)\propto
\nu^{-\beta}\,t^{-\alpha}$, where the spectral energy index is related
to the spectral photon index by $\Gamma=1+\beta$.  
%Uncertainties are quoted at $1\sigma$ confidence level.%, unless otherwise noted.
We employ standard cosmology with $H_{0}=71$ km s$^{-1}$ Mpc$^{-1}$,
$\Omega_{\Lambda}=0.73$, and $\Omega_{\rm M}=0.27$.
Quantities are listed in the cosmological rest-frame of the explosion unless explicitly noted.
%%%%%%%%%%%%%%%%%%%%%%%%%%%%%%%%%%%%%%%%%%%
\section{Sample selection and data analysis}
\label{Sec:Obs}

We study the late-time ($0.5<t<10$ days, rest-frame) X-ray emission of nearby ($z<0.5$)
long GRBs. 
At this epoch the X-rays are expected to be dominated by afterglow synchrotron emission
produced as the explosion shock is decelerated by the interaction with the circumburst medium,
whereas 
X-ray flares, steep decays and plateaus dominate at much earlier epochs (e.g. \citealt{Margutti13}). 

We select our sample of GRBs based on the following requirements: (i)  redshift measurement,
either from the optical afterglow or from unambiguous association to the host galaxy; (ii) bright early-time
($t<0.5$ days, rest-frame) X-ray emission to extract a spectrum and constrain the intrinsic 
neutral hydrogen absorption column $NH_{x,i}$; (iii) enough late-time ($0.5<t<10$ days, rest-frame) 
count statistics to constrain the X-ray spectral photon index $\Gamma_x$;
(iv) limited Galactic absorption along the line of sight $NH_{MW}\lesssim 10^{21}\,\rm{cm^{-2}}$
to reliably constrain $NH_{x,i}$ and avoid important contamination 
from Galactic material.\footnote{Existing maps of the projected spatial distribution of $NH_{MW}$ like those presented
in \cite{Kalberla05} are not able to capture possible variations of $NH_{\rm{MW}}$
on small angular scales.} Note that we do \emph{not} require the GRBs to have been
detected by the \emph{Swift} Burst Alert Telescope (BAT, \citealt{Barthelmy05}). What we require is an X-ray follow up of
the bursts both at early and at late times. At the time of writing, only  the \emph{Swift} X-ray Telescope (XRT, \citealt{Burrows05})
can provide early time X-ray follow up.

Furthermore, we require the GRBs to be at low redshift $z<0.5$ (i) 
to sample both  extremes of the distribution of $NH_{x,i}$ values (at higher redshift
the effective instrumental bandpass sensitive to intrinsic X-ray absorption decreases, thus reducing our 
ability to measure low $NH_{x,i}$ values, (e.g. \citealt{Margutti13}, their Fig. 5);
(ii) to minimize the effects of the detector sensitivity and band-pass when measuring
the GRB prompt emission duration $T_{\rm{90}}$ (e.g. \citealt{Littlejohns13}); 
%(the measured duration of a GRB generally decreases
%for increasing $z$, as a result of a poorer signal-to-noise ratio and intrinsic energy band shifting)(iii)
(iii) to sample the soft X-ray emission at $E\lesssim0.5\,\rm{keV}$ in the burst rest-frame.
This last requirement is essential to detect additional spectral components to the standard 
afterglow emission that could otherwise be missed in higher-$z$ GRBs if intrinsically soft.
An  example in this respect is the  black-body component with $kT_{BB}\sim0.5\,\rm{keV}$ 
that has been recently reported by \cite{Piro14} in the late-time X-ray emission of GRB\,130925A
at $z=0.347$.

Twelve GRBs satisfy the selection criteria above (Table \ref{Tab:data}). \emph{Swift}-XRT
data have been reduced as we describe in \cite{Margutti13}, using the latest software and
calibration files. Comparison with the online \emph{Swift}-XRT catalog (\citealt{Evans09}, \citealt{Evans10})
reveals a good agreement. For each GRB we 
measure the intrinsic absorption $NH_{x,i}$ from a spectrum extracted at $t<0.5$ days rest-frame, 
during a time interval where no spectral evolution is apparent. 
The Galactic contribution in the direction of the burst $NH_{MW}$ is estimated from \cite{Kalberla05}.
We constrain the late-time spectral photon index $\Gamma_x$ by extracting a spectrum in the time
interval $0.5<t<10$ days, rest-frame, while we estimate the power-law index of the temporal decay 
$\alpha_x$ by fitting the X-ray light-curve in the same time interval.  
All the spectra have been modeled with an absorbed power-law ($tbabs*ztbabs*pow$
within \emph{Xspec}). 
The duration of the burst prompt emission $T_{\rm{90}}$ is taken from \cite{Sakamoto11c},
\emph{Swift} Burst Alert Telescope (BAT) refined circulars or dedicated papers.

The results from this analysis are listed in Table \ref{Tab:data} and plotted in Fig. \ref{Fig:closure} 
and \ref{Fig:NHT90} (filled stars). We add for completeness 
the two pre-\emph{Swift} GRBs that would pass our selection criteria, 
GRBs 980425 and 030329 (open stars). Data have been collected from \cite{Tiengo03}, 
\cite{Kouveliotou04} and \cite{Kaneko07}. The following discussion will however focus on the 
\emph{Swift}-XRT sample, only, for the sake of homogeneity.

%%%%%%%%%%%%%%%%%%%%%%%%%%%%%%%%%%%%%%%%%%%
\section{Results}
\label{Sec:Res}

\begin{figure}
\vskip -0.0 true cm
\centering
\includegraphics[scale=0.60]{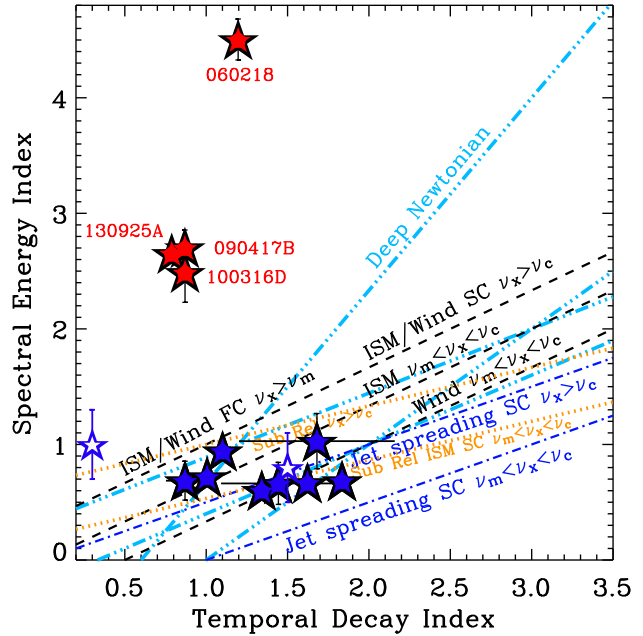}
\caption{Late-time (0.5-10 days, rest frame) spectral energy index $\beta_x$ vs. temporal decay index $\alpha_x$
for the sample of nearby GRBs (filled stars). Lines: expectations from synchrotron radiation from a relativistic, sub-relativistic
or newtonian shock expanding into an ISM or wind-like medium (see \citealt{Margutti13b} and references therein).
GRBs with evidence for super-soft X-ray emission are in red. Open stars: pre-\emph{Swift} GRBs
that satisfy the selection criteria of Sec. \ref{Sec:Obs}, i.e. GRBs 980425 and 030329.}
\label{Fig:closure}
\end{figure}

\begin{figure}
\vskip -0.0 true cm
\centering
\includegraphics[scale=0.55]{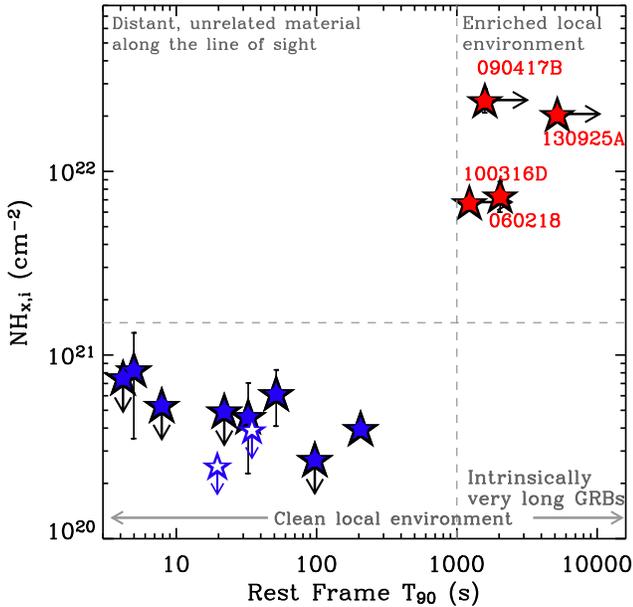}
\caption{Intrinsic hydrogen column density vs. prompt duration for the sample of nearby GRBs.
Red stars: GRBs with evidence for super-soft X-ray emission at late times. Open stars: pre-\emph{Swift} GRBs
that satisfy the selection criteria. The dashed horizontal line marks the peak of the $NH_{x,i}$ probability density distribution
of all GRBs detected by \emph{Swift} as of July 16th, 2014 \citep{Evans09}. The vertical dashed line at 1000 s
is drawn as a visual guide. 99\% of GRBs detected by \emph{Swift}-BAT show a rest-frame $T_{\rm{90}}<300$ s.}
\label{Fig:NHT90}
\end{figure}

\begin{deluxetable*}{llccccc}
\tablecaption{Prompt emission and late-time X-ray emission parameters}
\tablewidth{0pt}
%\tablewidth{\textwidth}
\tablehead{
\colhead{GRB} & z &\colhead{$T_{\rm{90, obs}}$} & \colhead{$NH_{x,i}$} 			       & \colhead{$\Gamma_x$}&$\alpha_x$& $NH_{MW}$\\
                            &    &\colhead{(s)}                              & \colhead{($10^{22}\,\rm{cm^{-2}}$)} &                                             & 		  &\colhead{($10^{22}\,\rm{cm^{-2}}$)} \\\
}
\startdata
060218 &  0.0331 & $2100\pm 100^{a}$		&$0.74\pm0.14$	&$5.5\pm0.2$		&$1.20\pm0.08$&0.140\\
060512 &  $0.443^{h}$   &$11.4\pm4.1$	&$<0.054$		&$2.0\pm0.2$		&$1.68\pm0.47$&0.016\\
060614 &  0.125   &$109.2\pm3.4$	&$<0.027$		&$1.69\pm0.04$	&$1.83\pm0.04$&0.020\\
061021 &  0.346   &$43.8\pm5.6$	&$0.046\pm0.024$	&$1.94\pm0.04$	&$1.10\pm0.03$&0.055\\
090417B & 0.345  &$>2130^{b}$		&$2.4\pm0.4$		&$3.7\pm0.2$		&$0.87\pm0.08$&0.017\\
091127    & 0.490  &$7.4\pm0.2$	&$0.084 \pm 0.05$	&$1.67\pm0.04$	&$1.62\pm0.04$&0.031\\
100316D & 0.0590&$>1300^{c}$		&$0.68\pm0.02^{d}$	&$3.5\pm0.3^{d}$		&$0.87\pm0.08$&0.101\\
120422A  & 0.283  &$5.4\pm 1.4$	&$<0.076$		&$1.7\pm0.2$		&$0.87\pm0.11$&0.042\\
130427A &  0.340  &$276.0\pm 5.0^{e}$&$0.040\pm0.004$	&$1.60\pm0.02$	&$1.34\pm0.01$&0.019\\
130702A &  0.145  & $59.0^{f}$		&$0.062\pm0.021$	&$1.72\pm0.03$	&$1.00\pm0.02$&0.018\\
130831A & 0.4791 &$32.5\pm2.5$	&$<0.050$		&$1.7\pm0.2$		&$1.44\pm0.35$&0.057\\
130925A  & 0.347   &$>7000$		&$2.05 \pm  0.20$	&$3.65\pm0.03$	&$0.79\pm0.02$&0.018\\
\enddata
\tablecomments{90\% c.l. uncertainties are provided for $NH_{x,i}$. All other uncertainties are
$1\,\sigma$. Upper limits are $3\sigma$.\\
$^a$ From \cite{Campana06}.\\
$^b$ From \cite{Holland10}.\\
$^c$ From  \cite{Starling11}.\\
$^d$ From \cite{Margutti13b}.\\
$^e$ From \cite{Maselli14}.\\
$^f$ From \cite{Collazzi13}, 50-300 keV band.\\
$^g$ This event has been classified as ultra-long, with a total duration of the prompt emission of $\sim20\,\rm{ks}$  
\citep{Evans14} and $\gamma$-rays lasting $\sim7\,\rm{ks}$ \citep{Piro14}.\\
$^h$ The redshift of this burst has been disputed by \cite{Fynbo09}. These authors suggest $z=2.1$ based on the assumed association of a broad absorption line in the afterglow spectrum of GRB\,061512 with $Ly\alpha$. Here we follow the analysis by \cite{Bloom06}. We note that our major conclusions are not sensitive to this particular choice.}
\label{Tab:data}
\end{deluxetable*}

Our analysis identifies a population of nearby GRBs with super-soft ($\Gamma_x>3$) 
X-ray emission at late-times (red stars in Fig.  \ref{Fig:closure} 
and \ref{Fig:NHT90} ), that is \emph{not} consistent with 
afterglow radiation from the forward shock, as shown in Fig. \ref{Fig:closure}.
Four bursts belong to this class: GRBs 060218, 090417B, 100316D and 130925A.
Their late-time temporal decay is also shallower than average: $\alpha_x<1.4$,
where $\alpha_x=1.4$ is the median value for the population of GRBs with known
redshift (\citealt{Margutti13b}, their Fig. 4). This finding alone is suggestive of the presence
of an additional X-ray emission component with markedly different spectral properties with
respect to the afterglow. Indeed, this was the conclusion from accurate broad-band spectral modeling
of the emission from GRBs 060218 (\citealt{Soderberg06c}), 090417B (\citealt{Holland10}),
100316D (\citealt{Margutti13b}) and, more recently, 130925A (\citealt{Evans14}, \citealt{Piro14}, 
\citealt{Bellm14}).

Remarkably, we find that the class of late-time super-soft X-ray emitters (red, filled stars) 
also shows significantly larger intrinsic absorption $NH_{x,i}$ \emph{and} 
exceptionally long duration of the prompt emission $T_{\rm{90}}$, as 
demonstrated in Fig. \ref{Fig:NHT90}. These bursts are also associated with fairly
large intrinsic optical extinction values: for GRBs 130925A and 090417B   \cite{Evans14} 
and \cite{Holland10} infer $A_V\approx 2.2$ mag and $A_V> 2.5$ mag, respectively.
In the case of GRB\,100316D, \cite{Olivares12} derive $A_V\approx 1.2$ mag,
while the intrinsic extinction along the line of sight to GRB\,060218 is likely lower
($E(B-V)_{\rm{tot}}=0.13\pm0.02$ mag, with a dominating Galactic component, \citealt{Pian06}).

In the $NH_{x,i}$-$T_{\rm{90}}$-$\Gamma_x$
phase space the 12 nearby GRBs naturally divide into two groups with no apparent continuum in 
between. The first group comprises GRBs with large intrinsic absorption 
$NH_{x,i}\gtrsim7\times 10^{21}\,\rm{cm^{-2}}$,  extremely long prompt emission  
$T_{\rm{90}}>1000$ s and super-soft late-time X-rays 
$\Gamma_x>3$. Low $NH_{x,i}<10^{21}\,\rm{cm^{-2}}$ is instead always associated with
a harder X-ray spectrum with $\Gamma_x\lesssim 2$ consistent with the predictions of the
afterglow model and a shorter $T_{\rm{90}}$. 
We estimate the probability 
to obtain the observed configuration by chance below.

Every GRB in our sample can be described by three stochastic variables. 
Each variable has two possible states, ``up'' or ``down'',
corresponding to large or small values of $NH_{x,i}$,
$T_{\rm{90}}$ and $\Gamma_x$, respectively. The
observed configuration corresponds to the case where every GRB
is either ``up-up-up'' or ``down-down-down'', implying that in our system of 12 GRBs
only 2 of the $2^{3}=8$ available states are populated. The chance probability
that in a system of $N$ elements only $n\leq2$ of the $m$ possible states are occupied 
is $P=\frac{(2^{N-1}-1)(m-1)+1}{m^{N-1}}$. For $N=12$ GRBs and $m=8$,
$P=1.7\times10^{-6}$. This calculation assumes equal probabilities for the ``up'' and ``down''
state of each variable ($p_{up}=p_{down}=0.5$), since a priori there is no reason to believe 
that any of the two states should be favored against the other. If however this is not true and 
the ``up'' state is intrinsically less probable with
$p_{up}=1/3$ ($p_{down}=2/3$) as suggested by the observations, then a Monte Carlo simulation
with $10^6$ realizations finds $P=1.3\times10^{-4}$. We note that for both calculations
we conservatively counted as ``success'' also the cases where $n=1$ (i.e. all the events in the same
state). % and cases where $m=2$
%but that would not be considered of any physical interest. 
We can thus reject the hypothesis of a chance association with high confidence ($P>99.99\%$).

We end by commenting on potential observational biases.  Any instrumental detection bias
would equally affect bursts in low and high $NH_{x,i}$ environments ($\gamma$-rays are not sensitive
to the $NH_{x,i}$) regardless of their late-time X-ray spectrum (the 
prompt emission ``knows'' nothing about the late-time X-ray spectrum). As a result  it could not
be responsible for the observed $NH_{x,i}$-$T_{\rm{90}}$-$\Gamma_x$ distribution.
Our sample is biased towards GRBs with brighter late-time X-ray emission, 
as we require enough count statistics to extract a spectrum at $0.5<t<10$ days (rest-frame): 
it is however unclear how this would only affect GRBs with short $T_{\rm{90}}$ and high $NH_{x,i}$
or long $T_{\rm{90}}$ and low $NH_{x,i}$. We therefore conclude that the observed configuration 
with two distinct clusters of GRBs 
in the $NH_{x,i}$-$T_{\rm{90}}$-$\Gamma_x$ phase space is physically driven.\\ 
%%%%%%%%%%%%%%%%%%%%%%%%%%%%%%%%%%%%%%%%%%%
\section{Discussion}
\label{Sec:Disc}

%-------------------------------------------------------
\subsection{The $\Gamma_x-T_{90}-NH_{x,i}$ phase space}

In the  collapsar model of long GRBs \citep{MacFadyen01} 
the duration of the prompt emission reflects
the central engine activity, with no expected connection with the amount of
circum-burst material or the spectral properties of the late-time X-ray emission. 
Our results indicate instead that nearby GRBs do \emph{not} populate the 
$NH_{x,i}$-$T_{\rm{90}}$-$\Gamma_x$
phase space randomly, and that these parameters
are in some way physically linked. In particular, it suggests that 
for the super-soft X-ray emitters the
measured $NH_{x,i}$ is dominated by material that is directly connected with the
explosion and/or progenitor, as opposed to more distant material that just happens to be along our line of sight.

The extreme X-ray softness of the late-time emission of some GRBs
has been noticed before (\citealt{Soderberg06c}, \citealt{Fan06b},
\citealt{Holland10}, \citealt{Margutti13b}, \citealt{Evans14}, \citealt{Piro14}, \citealt{Zhao14}, \citealt{Barniol14}). 
While there is general agreement on the need for an extra, super-soft X-ray component 
in addition to the afterglow, different ideas have been proposed to explain its origin.

(i) Radiation from a long-lived GRB central engine, later reprocessed by material in the
burst surroundings before reaching the observer (\citealt{Soderberg06c}, \citealt{Fan06b},
\citealt{Margutti13b}, see however \citealt{Barniol14}). 

(ii) Thermal emission from a hot 
($kT_{BB}\sim 0.5\,\rm{keV}$) and relatively compact $(R\sim10^{11}\,\rm{cm})$ cocooon 
that develops as a result of the interaction of the jet with the stellar layers. This model
was specifically developed for GRB\,130925A \citep{Piro14} but it is supposed to embrace
the entire class of GRBs with ultra-long prompt emission (see e.g. \citealt{Gendre13}
for details about ultra-long GRBs). 
This picture connects the late-time super-soft 
X-ray emission with the atypical nature of the progenitor star (a blue super-giant -BSG- instead of a Wolf-Rayet
star, whose outer layers power a longer-than-average jet activity, \citealt{Woosley12}), thus explaining the link  
between $\Gamma_x>3$ and $T_{90}>1000$ s. However, this model does not offer a natural explanation
for the large $NH_{x,i}$ (BSG progenitors actually have low mass-loss rates $\dot M<10^{-5}\,\rm{M_{\odot}yr^{-1}}$
even in super-solar metallicity environments, \citealt{Vink01}). 

(iii) Alternatively, a localized dust layer
at $R_d\sim 30-80\,\rm{pc}$  (GRB\,090417B, \citealt{Holland10}) or at $R_d\sim80-2000\,\rm{pc}$
(GRB\,130925A, \citealt{Evans14}, \citealt{Zhao14})
can account for both the spectral softness and the large intrinsic hydrogen column. However,
there is no reason to expect a longer than average duration of the prompt emission, contrary
to what is observed, \emph{if} the dust sheet is unrelated to the explosion/progenitor. 

The key question is how to connect the properties of the very early $\gamma$-ray emission,
late-time X-ray radiation and local environment density distribution within a coherent 
physical picture. GRBs that just happen to be seen through a 
thick but unrelated sheet of material would homogeneously populate the
upper part of the $NH_{x,i}$-$T_{90}$ plot (Fig. \ref{Fig:NHT90}), while
bursts where the central engine activity is entirely responsible for the duration of the
$\gamma$-ray emission are expected to reside in the lower part of the diagram, \emph{both} at short and
at long $T_{90}$. 

To explain the observed distribution of GRBs, 
with two well defined clusters in the $\Gamma_x-NH_{x,i}$-$T_{90}$ phase space,
we envision two scenarios:
 (i) the duration of \emph{all} the events is intrinsic (i.e. it
indeed reflects the duration of the activity of the GRB engine) and the longest events originate from
peculiar progenitors that only form in very dusty environments, 
or, more likely, (ii)  a single physical mechanism is responsible for the 
simultaneous appearance of the super-soft X-ray emission at late times, extremely long
$T_{90}$ and large $NH_{x,i}$. We expand on this latter possibility below.
We consider the first scenario less likely, as it would require a peculiar
progenitor with very large radius (to accommodate for the exceptionally long $T_{\rm{90}}$) 
to form in a peculiar dusty environment. However, we note that current 
observations do not allow us to rule out this possibility.\footnote{After our work appeared on the
public archive, the possibility of a larger progenitor star with a dense core engulfed in a low-mass extended
envelope has been suggested by \cite{Nakar15} to explain the $\gamma$-ray and UV properties of
GRB\,060218.}

%-------------------------------------------------------
\subsection{The role of progenitor mass loss in GRBs with late-time super-soft X-rays}

Wolf-Rayet stars (WRs) with $M\approx 40\,\rm{M_{\sun}}$ are considered the most likely progenitors of GRBs 
(e.g. \citealt{MacFadyen99}) and progenitor candidates of at least some ordinary
hydrogen-poor SNe (i.e. Type IIb, Ib and Ic SNe).
During the helium burning phase WRs lose mass through powerful winds at the
rate of $\dot M\sim 10^{-5}\,\rm{M_{\sun}yr^{-1}}$ with velocity $v_w\sim1000\,\rm{km\,s^{-1}}$. 
However, during the last $\sim100-1000$ yrs,
evolved WRs burn heavier elements and the mass-loss rate is \emph{not} well constrained. 
As a result, the mass distribution within $\sim10^{17}-10^{18}\,\rm{cm}$ is unknown and
might strongly deviate from the $\propto1/r^2$ wind profile. Three recent observational 
findings are relevant in this respect:
(i) the possible eruption of the Type Ic SN PTF11qcj $\sim2$ yrs before the 
supernova \citep{Corsi14}; (ii) the indication of increased mass loss with clear WR signatures
shortly before the explosion in the very early spectra of the envelope-stripped SN2013cu (\citealt{GalYam14},
see also \citealt{Groh14}) and the inference of an increased mass-loss in the days before the explosion
of SN\,2008D \citep{Svirski14}; (iii) the detection of modulated radio emission from nearby envelope-stripped SNe,
which is indicative of pre-explosion mass loss variability (\citealt{Soderberg06d}, \citealt{Wellons12}).

%Progenitor Winds
%Massive Wolf-Rayet stars -during helium burning- are known to have large mass loss rates, approximately
%$10^{-5}$ solar masses/yr or more. This wind may be clumpy and anisotropic and its
%metallicity dependence is uncertain. The density dependence of
%matter around a single star in vacuum could be approximately $1000 (10^{16} cm/R)^2$ cm-3
%composed of carbon, oxygen, and helium,
%BUT During approximately the last
%$\sim100 - 1000$ years of its life, the star burns carbon (mainly) and heavier fuels.
%The mass loss rate of the star during these stages is
%unknown. No WR star has ever knowingly been observed
%in such a state. This means that the mass distribution
%inside $\sim10^{17-18}$ cm is unknown (100 yrs at 1000 km/s).

We suggest that the main difference between the two groups of GRBs in Fig. \ref{Fig:closure} and
\ref{Fig:NHT90} is connected with the distribution and amount of material in the burst
local environment and that the interaction of the explosion's shock and 
radiation with the medium enriched by substantial mass loss
from the progenitor star is the source of the phenomenology observed in GRBs with super-soft X-ray emission.

In particular, \emph{we associate the late-time super-soft X-ray emission with reprocessed radiation
by material in the burst surroundings}. A possibility is a dust echo of the prompt X-rays
as it was suggested for GRBs 090417B and 130925A (\citealt{Holland10}; \citealt{Evans14}).
Following  \cite{Shao08} and \cite{Shen09} (their Eq. 4)  and noting that the X-ray ``plateau'' emission decays around 
$(1-5)\times10^4\,\rm{s}$ in the four GRBs with $\Gamma_x>3$, we find that the dust sheet is 
roughly located at  $R\sim$ ten to a hundred pc from the progenitor. 
%Alternatively, a Compton echo of the prompt $\gamma$-rays is also possible \citep{Madau00}. 
%Comptonization of the hard photons as they propagate
%through the medium to the observer would naturally suppress the high-energy radiation,
%leading to a spectral cutoff at $\sim 511/\tau_s^2\,\,\rm{keV}$.
%The observed very soft spectrum would require an optical depth $\tau_s\sim 10-15$.
Differently from previous works, here we argue that the physical origin of the material
around the GRBs with $\Gamma_x>3$ has to be connected with the mass-loss history of the progenitor system.  

The variable mass loss from massive stars during their evolution is known to create
a structured wind bubble around the star (and its possible binary companion)
with a  dense shell at the interface with the ISM (e.g. \citealt{Moore00}).
The typical radius of the swept-up shell nebula around a WR star at the end of its life is 
$R\approx 100(\dot M_{-6.2}v^{2}_{3.5}n_{0}^{-1})^{1/5}t_{6.6}^{3/5}$ pc, with critical 
dependency on the main sequence (MS) mass-loss rate $\dot M$, wind velocity $v$,
MS lifetime $t$ and ISM density $n$ \citep{Castor1975}. Here we normalize our
variables to $\dot M=10^{-6.2}\,\rm{M_{\sun}yr^{-1}}$,
$v=10^{3.5}\,\rm{km\,s^{-1}}$, $t=10^{6.6}$ yr and $n=1\,\rm{cm^{-3}}$, as appropriate
for a $35\,M_{\sun}$ star (e.g. \citealt{Dwarkadas07}). Observations show that 
shell nebulae are present in $\sim35\%$ of WRs \citep{Marston97}
with $R\sim$ tens of pc, intriguingly similar to what we infer for the light-echo scenario above. 
%More to this point, dust shells with $R\sim$ have been observed around the WC 
%variety of WR stars, which are a final evolutionary phase of WRs. 
%Binarity appears to play a key role in the dust formation of these systems \citep{Crowther07}.
We suggest that 
the excess of super-soft X-ray radiation at late times is related with the 
presence of shell nebulae around the progenitors.\footnote{We emphasize that at this stage it is not possible
to rule out the possibility that the excess of super-soft X-rays
originates from radiation from a long-lived central engine, later reprocessed by material 
in the burst environment before reaching the observer.
However, the ``echo hypothesis'' seems more natural as it does not require
an extremely long central engine life-time of several days.}

%strong rotation leads to the formation of a wind
%that is faster and denser at the poles of the star as compared to
%the equator (Maeder & Desjacques 2001; Dwarkadas & Owocki
%2002).

The interior structure of the wind bubble is instead determined by the 
more recent mass-loss history of the progenitor and possibly results
from the ejection of massive shells of material that shaped the medium at $R<1$ pc.
We speculate that the interaction of the explosion's jet/shock with material 
at $R\sim10^{14}-10^{16}\,\rm{cm}$ is at least partially responsible for the very long 
prompt duration ($T_{90}>1000$ s) in these sources. 
%(external shock model for the prompt by Dermer). 
In the most extreme cases,
(i.e. weak or absent explosion's jet and/or thick shell of material in the very close environment) 
the entire $\gamma$-ray emission originates from
shock/jet break out radiation through a thick shell, as it was proposed for GRBs 060218 and 100316D
by \cite{Nakar12} (see also \citealt{Bromberg11}). 
In this model the $T_{90}$ reflects the properties of the progenitor star -and in particular its
radius- or its environment, as opposed to the duration of the central engine activity.
For GRBs 060218 and 100316D, \cite{Nakar12} obtained a break-out radius 
$R_b\sim5\times 10^{13}\,\rm{cm}$, much larger than WR stars, and pointed to the 
presence of opaque material  thrown by the star before exploding at $R=R_b$.

%-------------------------------------------------------
\subsection{Testable predictions}

\emph{If} the interaction of the jet with some material is at least partially 
responsible for the exceptionally long $T_{90}$, we expect that: 
(i) the interaction %with material left by the progenitor 
would suppress the shortest variability timescales originally present; 
(ii) we expect the GRBs with the hardest prompt emission to belong to the naked-explosion
category, and the softest events to belong to the super-soft X-ray emitters, with 
overlap between the two classes due to the large variation in the intrinsic properties 
of the explosions (i.e. before the interaction with the medium). 

Observations confirm these expectations. GRBs 060218 and 100316D are characterized by 
very low prompt peak energy values
$E_{pk}<50\,\rm{keV}$ (\citealt{Kaneko07}, \citealt{Starling11})\footnote{For completeness we
report here that the different episodes of emission of GRB\,130925A have $E_{pk}=60-100\,\rm{keV}$
as measured from a Band spectrum by \cite{Evans14}. GRB\,090417B instead lacks a high-energy 
follow up at $E>150\,\rm{keV}$ and its spectrum, as detected by the \emph{Swift}-BAT, is consistent
with a simple power-law with spectral photon index $\Gamma=1.9\pm0.1$ \citep{Holland10}.} and show $\Gamma_x>3$ at late times, while
GRB130427A, with peak energy $E_{pk}\sim1400\,\rm{keV}$ during the main emission episode
(\citealt{Maselli14}), belongs to the naked-explosion category
and it is the hardest long GRB at $z\lesssim 0.3$. Furthermore, the very smooth, single-peaked 
temporal structure of  the prompt emission of  GRBs 060218 and 100316D is clearly in line with the 
shock break-out scenario \citep{Nakar12}, and it is apparent from Fig. 1 of
\cite{Holland10} that for GRB\,090417B the shortest variability timescales  were not observed
($\delta t_{var}> 10\,\rm{s}$). Finally, it is remarkable that
 in spite of the excellent statistics GRB\,130925A also shows
a large minimum variability time-scale $\delta t_{var}^{min}\sim1\,\rm{s}$ (\citealt{Greiner14}) with typical values
$\delta t_{var}\sim10-100\,\rm{s}$ (\citealt{Evans14}, Fig. 5).  This finding points to  a large dissipation
radius $R=2\times 10^{14}(\delta t/1 \,\rm{s})(\Gamma/30)^2\,\rm{cm}$, consistent with the picture
that we propose.

We conclude with two comments. First, at the moment we 
have \emph{no} reasons to believe that no GRB will ever populate 
the lower-right and upper-left corners of the $NH_{x,i}$ vs. $T_{90}$ plot (Fig. \ref{Fig:NHT90}).
Instead, we expect chance alignment with thick but unrelated material to happen for
a certain fraction of GRBs. At the same time, we cannot exclude the existence of intrinsically
very long GRBs exploding in very clean environments in the low-redshift Universe. 
Since there is no obvious observational bias against
the detection of these two groups of explosions, we conclude that they must be less common.
Future observations are needed to clarify how GRBs populate the $NH_{x,i}$ vs. $T_{90}$ plane.
Second, the finding of low ambient densities around
some of the super-soft X-ray emitters as inferred from broad-band modeling of the afterglow 
emission (see e.g. GRB\,130925A, \citealt{Evans14})
is not in contrast with our picture of an environment with a complex density profile, as the general structure
around a massive star is that of a low density medium surrounded by an overdense shell.
Furthermore, repeated shell ejections might as well have swept up the material around the
progenitor, leaving behind a low-density cavity (in strict analogy with nova shells,
see e.g. Fig. 5 in \citealt{Margutti14b}).

%the late-time super-soft emitters divide into 2 pairs with different properties
%this solution is contrived and we conclude that ... is most likely...
%%%%%%%%%%%%%%%%%%%%%%%%%%%%%%%%%%%%%%%%%%%
\section{Conclusions}
\label{Sec:Conc}

We have identified a class of GRBs in the low-redshift Universe with (i) super-soft late-time X-ray radiation 
not consistent with the standard afterglow model, (ii) large X-ray absorption and (iii) exceptionally long
prompt $\gamma$-ray emission duration. We connect these properties with the
turbulent mass-loss history of their stellar progenitors that shaped the environment around the
explosions. We suggest that the interaction of the explosion's shock/jet and of the emitted radiation with the 
complex medium is responsible for the anomalous late-time X-ray spectrum and the extremely long duration of the
prompt emission in these sources. 

The next step in the research is to understand which 
physical property distinguishes these stellar progenitors from those giving origin to 
naked explosions. While this is at the moment unclear, we note that the ``peculiarity''
of the progenitors of this class of explosions might manifest
through different properties of their host-galaxies and, especially, of their local environments 
(e.g. metallicity at the explosion site, star formation rate, 
dust content, properties of the underlying stellar population).
Furthermore, we predict that the remnants of these explosions will be
significantly different from the cases of expansion into an undisturbed ISM or wind-like
density profile as the more complex structure of the medium will cause a number
of reflected and transmitted shocks to go through the explosion's ejecta
and the material in the burst surroundings. 
%\textbf{Here we need something related to remnants:
%enrico's paper  or Lopez papers or ref in Dwarkadas paper}

It is beyond current capabilities to spatially resolve the remnants of these explosions
due to their large distance. Additionally, we will not be able to witness their evolution in
real time: the supernova ejecta will interact with the more distant shell on a
timescale of $>300$ yrs, and even a powerful jet with $E=10^{52}\,\rm{erg}$
propagating in a wind medium with $\dot M=10^{-5}\,\rm{M_{\sun}yr^{-1}}$ would 
need several decades to reach $R=10$ pc. 
We will thus need to rely on the increased capabilities of current and near-future
optical surveys to localize bursts from their optical afterglow (as it indeed happened
for GRB\,130702A, \citealt{Singer13}) to build the statistical sample of GRBs in the low-redshift Universe necessary to 
understand how they populate the $NH_{x,i}$ vs. $T_{90}$ plane, 
constrain and contrast the properties of their environments, and test our expectations.

\acknowledgments 
We thank the referee for helpful comments that 
improved the quality of our work.
R.M. is grateful to the Aspen Center for Physics and the NSF Grant \#1066293 for hospitality 
during the completion of this work and for providing a stimulating environment that inspired
this project. Support for this work was provided by the David and Lucile Packard Foundation Fellowship
for Science and Engineering awarded to A.~M.~S.

%This manuscript was completed during the ``Fast and Furious:
%Understanding Exotic Astrophysical Transients'' workshop at the Aspen
%Center for Physics, which is supported in part by the NSF under grant
%No.\ PHYS-1066293.

\bibliographystyle{apj}

\end{document}